\documentstyle[preprint,aps,prb]{revtex}
\begin{document}
\draft
\title{ 
       Transport through a finite  Hubbard chain \\
       connected to reservoirs
}
\author{Akira Oguri}
\address{
          Department of Material Science,
          Faculty of Science,
          Osaka City University, \\
          Sumiyoshi-ku, Osaka 558-8585,
          Japan
}

\date{October 15, 1998}
\maketitle

\begin{abstract}
The dc conductance, $g_N$, through a finite Hubbard chain 
of size $N$ ($=1,\,2,\,3,\,\ldots$) connected 
to reservoirs is studied at $T=0$ in an electron-hole symmetric case. 
We calculate a spatial dependence  
of the self energy analytically at $\omega=0$ 
within the second order in $U$, 
and obtain an inter-site Green's function $G_{N1}$, 
from which $g_N$ can be determined, via the Dyson equation. 
The results depend strongly on whether $N$ is even or odd. 
For odd $N$, a perfect transmission occurs, 
and $g_{N} \equiv 2e^2/h$ independent of the values of $U$. 
This may be attributed to a Kondo resonance appearing at $\omega=0$. 
For even $N$, $g_{N}$ decreases with increasing $N$, 
and converges to a finite constant 
which is a smooth decreasing function of $U$.
These behaviors are essentially owing to the presence of the reservoirs, 
which make a quasi-particle description valid 
for low-energy states at $\omega \ll \hbar v_F/(Na)$; 
where $v_F$ is the Fermi velocity and $a$ is the lattice constant.

\end{abstract}


\narrowtext

Effects of an electron-electron interaction $U$
on the transport through a small system, 
such as a quantum dot and a quantum wire, 
have been a subject of current interest. 
So far, theoretical investigations for a quantum dot\cite{NL,GR}
and a quantum wire\cite{KaneFisher} seems to be developed separately 
because of the difference in their dimensionality.
The purpose of this work is to study a system of intermediate size 
and investigate how the transport depends on the size.
To this end, we consider a finite Hubbard chain of size $N$ 
connected to Fermi-liquid reservoirs  
which are simulated by semi-infinite noninteracting leads 
(see Fig.\ \ref{fig:lattice}),
and calculate the dc conductance at $T=0$ 
with the perturbation approach\cite{YY1-2}
starting from a zero-dimensional limit.
For small $N$, 
the system we consider may be regarded as a model for a quantum dot 
consisting of multi-levels, 
which has recently been examined by advanced numerical methods, 
such as a numerical renormalization group\cite{Izumida3} 
and a quantum Monte Carlo method.\cite{ao6} 
Specifically, in the present study, 
we concentrate on an electron-hole symmetric case. 
Thus, for large $N$, the system may be regarded as a model 
for a Mott-Hubbard insulator, 
which has been studied by a number of groups 
with the bosonization approach 
\cite{PonomarenkoNagaosa,OdintsovTokura,StarykhMaslov,MoriOgataFukuyama,FujimotoKawakami} 
starting from a one-dimensional limit.

We now proceed to details.  
The model consists of three regions (see Fig.\ \ref{fig:lattice});
a finite Hubbard chain at sites $\,1 \leq i \leq N$, 
and two semi-infinite leads on the left $\,-\infty < i \leq 0$, 
and the right $\,N+1 \leq i < +\infty$. 
The Hamiltonian is given by
\begin{eqnarray}
   {\cal H} \ 
    &=&  \ {\cal H}_L + {\cal H}_R  
  + {\cal H}_C^0 + {\cal H}_C^{int}
+ {\cal H}_{mix} 
\label{eq:H}
\;, \\
{\cal H}_L &=& 
        \sum_{\scriptstyle i,j=-\infty \atop \scriptstyle \sigma}^{0} 
        \left[\,- t_{ij}  -\mu\, \delta_{ij} \, \right]
            c^{\dagger}_{i \sigma} c^{\phantom{\dagger}}_{j \sigma}
\;,
\\
{\cal H}_R &=&  
        \sum_{\scriptstyle i,j=N+1 \atop \scriptstyle \sigma}^{\infty} 
        \left[\,- t_{ij}  -\mu\, \delta_{ij} \, \right]
             c^{\dagger}_{i \sigma} c^{\phantom{\dagger}}_{j \sigma}
  \;, 
\\
  {\cal H}_C^0 &=&  
     \sum_{\scriptstyle i,j=1  \atop \scriptstyle \sigma}^{N} 
        \left[\,- t_{ij}  
   \, + \, 
\left(\, \epsilon_0 + {U \over 2} - \mu \,\right) \delta_{ij}\,\right]
             c^{\dagger}_{i \sigma} c^{\phantom{\dagger}}_{j \sigma}
\;, 
\\
  {\cal H}_C^{int} &=&   
U \,\sum_{i=1}^N
        \left[\,  n_{i \uparrow}\,n_{i \downarrow} 
         - {1\over 2}\, (n_{i \uparrow} + n_{i \downarrow} )
        \,\right]
\;,
\\
  {\cal H}_{mix} &=& 
- v_L\, \sum_{\sigma} \left(\,  
             c^{\dagger}_{1 \sigma} c^{\phantom{\dagger}}_{0 \sigma}
             + \mbox{H.c.} \,\right) 
- v_R \, \sum_{\sigma}  \left(\,  
             c^{\dagger}_{N+1 \sigma} c^{\phantom{\dagger}}_{N \sigma}
             + \mbox{H.c.} \,\right) 
  \;. 
\end{eqnarray}
Here $c^{\dagger}_{j \sigma}$ 
 ($c^{\phantom{\dagger}}_{j \sigma}$) is the creation (annihilation) 
operator for an electron with spin $\sigma$ at site $j$, 
$n_{j \sigma}^{\phantom{\dagger}} 
= c^{\dagger}_{j \sigma} c^{\phantom{\dagger}}_{j \sigma}$, 
$\mu$ is the chemical potential, $\epsilon_0$ is the on-site energy.
In each of the regions, the intra-region hopping matrix element $-t_{ij}$
is finite only for the nearest-neighboring sites: 
$-t_{ij} = -t\,\left[\, \delta_{i+1, j} + \delta_{i-1, j} \,\right]$.
The connection between the interacting region and the left (right) lead 
is described by the tunneling matrix element $v_L$ ($v_R$).
In the present study,  we treat $\,{\cal H}_C^{int}$ as a perturbation 
taking the unperturbed Hamiltonian to be 
$\,
{\cal H}^{(0)}  
    =   {\cal H}_L + {\cal H}_R  
  + {\cal H}_C^0 + {\cal H}_{mix} 
$. 
Especially, in a limiting case $N=1$, 
our model corresponds to a single Anderson impurity, 
in which the perturbation expansion is valid for all values of $U$.
\cite{ZlaticHorvatic}
Our working hypothesis is that 
the ground state is changed continuously
when $N$ is increased to intermediate size
owing to the presence of the noninteracting-leads.
\cite{ao6,ao5}

We next consider the retarded Green's function;  
\begin{equation} 
G_{jj'}(\omega+i0^+) 
 =  
- i   \int_0^{\infty} \! dt' \,
   \left \langle  \left\{   
   c^{\phantom{\dagger}}_{j \sigma} (t') ,\,  c^{\dagger}_{j' \sigma} (0)  
                  \right\}  \right \rangle  \, e^{i\, (\omega +i0^+) t'} 
\;,
\label{eq:G_Matsubara}
\end{equation} 
where $c_{j \sigma}(t') = 
       e^{i  {\cal H} t'} c_{j \sigma} e^{- i  {\cal H} t'}$,
$\,\langle \cdots \rangle\,$ denotes the average 
$\mbox{Tr} \left[ \, e^{- \beta {\cal H} }\, {\cdots}
\,\right]/\mbox{Tr} \, e^{- \beta {\cal H}  }$, $\beta=1/T$, and
the curly brackets denote the anticommutator.  
The spin index for the Green function has been omitted 
assuming the average to be independent of whether spin is up or down.
Since the interaction is restricted for the electrons in the central region,
the Dyson equation is written as
\begin{equation} 
  G_{ij}(z)    =   G^{(0)}_{ij}(z) 
    + \sum_{l,m=1}^{N}\,G^{(0)}_{il}(z)\,  \Sigma_{lm}(z)
   \, G_{mj}(z) \;,
  \label{eq:Dyson}
\end{equation} 
where $G^{(0)}_{ij}$ is the free Green's function 
corresponding to ${\cal H}^{(0)}$, and 
$\Sigma_{ij}(z)$ is the self-energy correction due to ${\cal H}_C^{int}$. 
Note that $G_{ij}  =  G_{ji}$ and $\Sigma_{ij} = \Sigma_{ji}$ 
because of the time reversal symmetry of ${\cal H}$. 
At $T=0$, the dc conductance is expressed in terms of 
$G_{N 1}(\omega+i0^+)$; \cite{ao6,ao5} 
\begin{equation}
g_N   =  {2 e^2 \over h} \ 
        4\, \Gamma_L(0)\, \Gamma_R(0) \,  \left| G_{N 1}(i0^+) \right|^2 
\;,  
\label{eq:cond}
\end{equation}
where $\,\Gamma_{\alpha}(0) =  \pi \, D(0)\, v_{\alpha}^2$ 
with $\,\alpha=L,R\,$ and $\,D(0)=\sqrt{4t^2-\mu^2}\,/\,(2 \pi t^2)$ 
being the local density of states at an edge of the isolated lead. 
The Green's function for $\omega=0$, $G_{N 1}(i0^+)$, 
can be calculated solving the Dyson equation Eq.\ (\ref{eq:Dyson})
taking the self-energy $\Sigma_{ij}(i0^+)$ as an input.
This part of the calculation can be regarded as  
a scattering problem of a free quasi-particle.\cite{HewsonX}    
It is described by an effective Hamiltonian
$\,\widetilde{\cal H}^{(0)}_{qp}   \equiv   {\cal H}_L + {\cal H}_R 
                         + \widetilde{\cal H}_C^0 + {\cal H}_{mix}$,
which has a renormalized matrix element,
$-\widetilde{t}_{ij}^C  \equiv  -t_{ij}  +  \mbox{Re}\,\Sigma_{ij}(0)$, 
in the central region 
$\,
\widetilde{\cal H}_C^0  \equiv    
 \sum_{\scriptstyle i,j=1 \atop \scriptstyle \sigma}^N 
         \left[- \widetilde{t}_{ij}^C  
     +  
 \left( \epsilon_0 + U/2 - \mu \right) \delta_{ij}\right]
 c^{\dagger}_{i \sigma} c^{\phantom{\dagger}}_{j \sigma} 
$.
This mapping onto the free quasi-particle is justified under the condition 
$\,\mbox{Im}\,\Sigma_{ij} (i0^+)=0$ at $T=0$,
which can be shown with the perturbation theory.\cite{LangerAmbegaokar}
In other words, when the ground state is a Fermi-liquid,
there exists a one-particle Hamiltonian which reproduces 
the Green's function exactly in the limit $\omega=0$, $T=0$.
Thus, the physical quantities which are expressed in terms of $G_{ij}(i0^+)$,
such as $g_N$ and the number of displaced electrons,\cite{LangerAmbegaokar}
are reproduced exactly by using $\widetilde{\cal H}^{(0)}_{qp}$.

We now calculate the self-energy.
In what follows, we take the tunneling matrix element to be $v_R= v_L =t$, 
for simplicity, and consider an electron-hole symmetric case,  
where the average number of electrons per site is unity, 
taking the parameters to be $\epsilon_0 + U/2 -\mu=0$ and $\mu=0$. 
In this case, the unperturbed Hamiltonian has a simple form 
${\cal H}^{(0)} = 
        -t\, \sum_{\scriptstyle i=-\infty, \sigma}^{\infty} 
        \left(\,
            c^{\dagger}_{i+1 \sigma} c^{\phantom{\dagger}}_{i \sigma}
  + \mbox{H.c.} \,\right)$,
and effects of $U$ starts from a second-order contribution 
illustrated in Fig.\ \ref{fig:diagram}. 
The corresponding retarded self-energy is written as
\begin{eqnarray}
\Sigma^{(2)}_{jj'} (\omega +i0^+) 
 &=&  U^2 
\int_{-\pi}^{\pi} {dk_3 dk_2 dk_1 \over (2\pi)^3 }\, 
e^{i\, (k_1 + k_2 -k_3) (j-j')}
 \nonumber \\
& &  \qquad
\times \  
{ 
               f_{k_3}(1-f_{k_2})(1-f_{k_1}) + (1-f_{k_3})f_{k_2}f_{k_1}  
               \over
               \omega + \epsilon_{k_3} -\epsilon_{k_2} -\epsilon_{k_1} +i0^+
             } \; .
\label{eq:Self_2}
\end{eqnarray}
Here $\,j$ and $\,j'$ are restricted 
in the central region $\,1\leq j, j' \leq L$, 
otherwise $\,\Sigma^{(2)}_{jj'}(\omega +i0^+) \equiv 0$,
$f_k=[ e^{\beta\epsilon_k} + 1 ]^{-1}$, and 
$\epsilon_k = -2t \cos k$.   
It is straight forward to show that
the imaginary part behaves as 
$\, -\mbox{Im}\, \Sigma^{(2)}_{jj'} (\omega +i0^+) \propto \omega^2\,$  
for small $\omega$ at $T=0$.
This behavior is quite different from 
that in a usual one-dimensional system, 
in which the imaginary part for a low-energy excitation 
at the Fermi point $k_F$ is proportional to $|\omega|$ 
as a result of the momentum conservation 
caused by the translational invariance.
We next consider the real part at $\omega=0$, $T=0$.
Because of the electron-hole symmetry, 
the real part becomes zero when  $|j-j'|$ is an even number ($=2m$);   
$\mbox{Re}\,\Sigma_{2m}^{(2)}(0) \equiv 0$.
On the other hand, when $|j-j'|$ is an odd number ($=2m+1$),
Eq.\ (\ref{eq:Self_2}) can be rewritten as; 
$\mbox{Re}\, \Sigma^{(2)}_{2m+1} (0)   
  = -t\, \left\{\rho(0)  U  \right\}^2  S_{2m+1}$ 
with  
$\,\rho(0) \equiv 1/(2\pi t)$ being  
the density of states corresponding to ${\cal H}^{(0)}$,  
and 
\begin{eqnarray}
 S_{2m+1} &\equiv& {1\over 2\pi}\, 
 \mbox{P} \int_{-\pi/2}^{\pi/2} dk_3 dk_2 dk_1 \, 
 { 
              \cos[k_1 (2m+1) ] \,
              \cos[k_2 (2m+1) ] \,
              \cos[k_3 (2m+1) ] 
                \over
              \cos k_1 \,  + \cos k_2 \, + \cos k_3 
             } 
\label{eq:Self_Hubbard_def}
\;.
\end{eqnarray}
Further, this expression can be simplified as
\begin{eqnarray}
 S_{2m+1} 
&=&  
 {(-1)^m\,\pi^2 \over 2} \, \int_0^{\infty} dz \, J_{2m+1} (z) 
      \left [\, 3 \left\{ \mbox{\bf E}_{2m+1}(z)\right\}^2  
                 - \left\{ J_{2m+1}(z)\right\}^2  \,\right] \;.
\label{eq:Self_Hubbard}
\end{eqnarray}
Here $J_{2m+1}(z)$ and $\mbox{\bf E}_{2m+1}(z)$
are the Bessel and Weber functions.\cite{AbramowitzStegun}
The integration can be performed analytically by decomposing 
it into two parts;\cite{ao7} 
\begin{eqnarray}
 \int_0^{\infty} dz \, \left\{J_{2m+1}(z) \right\}^3 
&=&{1 \over (2m+1) \pi} \, _3F_2
\left({1\over 2}, -2m- {1\over 2}, 2m+{3\over 2}; {1\over2}-m,m+ {3\over 2}; 
{1\over 4} \right)
\;,
\end{eqnarray}
where $_3F_2(a_1,a_2, a_3; b_1,b_2;\,z)$  
is the generalized hypergeometric function,\cite{Mathematica} 
and
\begin{eqnarray}
& &  \int_0^{\infty} dz \, J_{2m+1} (z) 
\left[\, 
\left\{J_{2m+1}(z) \right\}^2  
+ 
\left\{ \mbox{\bf E}_{2m+1}(z) \right\}^2 
\, \right] 
\nonumber \\
& & = \  
 \left({2 \over \pi}\right)^2 \,
\sum_{l=0}^m \, 
(-1)^l\,{m+l \choose 2l}
\nonumber \\
& & \ \times \Biggl\{\ 
   {2l \choose l}^2 \,{1 \over (2m+1)^2 } 
\ + \ {2l \choose l}
\sum_{k=1}^{l}  {2l \choose l-k}\,
{
2 \, (2m+1)^2 \, 
\left[\, 1 + (-1)^k \,\right] 
\over  \left[\,(2m+1)^2 -k^2\,\right]^2
}
\nonumber\\
& & \ \ \ \ + 
\ \sum_{k=1}^{l}
\sum_{q=1}^{l} 
 {2l \choose l-k}  {2l \choose l-q}\,
{
2 \, (2m+1)^2 \,
\left[\, (-1)^k + (-1)^q \,\right]  
\over \left[\,(2m+1)^2 -(k+q)^2)\,\right]\,  
\left[\,(2m+1)^2 -(k-q)^2)\,\right]
}
\ \Biggr\}
\;.
\end{eqnarray}
The result for $S_{2m+1}$ is summarized in the Table \ref{table1},
and is also plotted in Fig.\ \ref{fig:self2}.
The self-energy correction, $S_{2m+1}$, is largest 
for the nearest-neighboring sites $\,m=0$, 
and decreases rapidly with increasing $m$ 
showing an oscillatory behavior.

Using the above results for the self-energy, 
we determined the renormalized matrix element $-\widetilde{t}_{ij}^C$ 
within the second order, and calculated $G_{N1}(i0^+)$  solving 
the scattering problem of a free quasi-particle 
described by $\widetilde{\cal H}^{(0)}_{qp}$. 
Then, we obtained $g_N$ from Eq.\ (\ref{eq:cond}). 
In Fig.\ \ref{fig:g_N}, 
$g_N$ is plotted as a function of the size $N$ 
for $\rho(0) U=1.0$. 
When $N$ is an odd number ($=2M+1$), 
the perfect transmission occurs, $g_{2M+1} \equiv 2e^2/h$,
independent of the values of $U$. 
This remains valid even when the higher-order terms are included, 
which can be proved based on the quasi-particle description,\cite{ao7} 
and the perfect transmission is understood as a result of the electron-hole 
and the inversion ($v_L=v_R$) symmetries.
Physically, it may be attributed to a Kondo resonance 
appearing at $\omega=0$. 
In other words, this kind of an even-odd property is 
related to the level-structure of an isolated Hubbard chain of size $N$.
For an odd $N$, the isolated chain has two eigenstates at $\omega=0$, 
which corresponds to the Kramers doublet states. 
When the reservoirs are connected, 
the doublet states are changed to a Kondo resonance state and contribute 
to the perfect transmission. 
On the other hand, for an even $N$, 
there is no discrete eigenstate at $\omega=0$. 
These features reflect strongly on the transport 
of the connected system.

When $N$ is an even number, the dc conductance shown in Fig.\ \ref{fig:g_N}  
decreases with increasing $N$, 
and tends to a finite constant for large $N$.
In Fig.\ \ref{fig:g_2M} the dc conductance for the even case with $N=2M$ 
is plotted as a function of $M$ for several values of $\rho(0) U$ 
($= 0.5,\, 1.0,\,1.5$, and $2.0$). The value of
$g_{2M}$ converges well to a constant already at small $M$, 
and the constant decreases monotonically with increasing $U$. 
Figure \ref{fig:g_50} shows the $U$ dependence of $g_N$ for $N=50$. 
Within the accuracy of the line thickness, 
the curve can be regarded as an extrapolated value 
owing to the early convergence. 
We note that the reduction of $g_{2M}$ from 
the universal value $2e^2/h$ is the effects due to the electron correlation,
and within the Hartree-Fock approximation $g_{2M}$ is unchanged 
from the universal value.
In addition, our result shows a metallic behavior, 
{\em i.e.}, $g_{2M}$ is finite in the limit $M \to \infty$, 
in spite of the half-filled case. 
We have obtained this behavior 
by taking the limit $T \to 0$ first, keeping $M$ to be finite. 
In this limit, the sample and two leads are described 
by a single (ground state) wave function,
and the phase coherence among the whole system 
plays an important role for the transport property. 
If the limit $M \to \infty$ is taken first at finite temperature,
the phase coherence may be disturbed by thermal excitations.
Qualitatively, our result is consistent with 
that obtained by Ponomarenko and Nagaosa,\cite{PonomarenkoNagaosa}
and by Odintsov, Tokura and Tarucha\cite{OdintsovTokura}
with the bosonization approach.
Their results also show a metallic behavior 
bellow a characteristic temperature,
and effects of a Mott-Hubbard gap on the dc conductance 
appear above the characteristic temperature.

In the present study, we have studied a finite system connected 
to reservoirs described by the semi-infinite noninteracting leads. 
The presence of the reservoirs seems to make 
the quasi-particle description of a Fermi-liquid  
valid at lower temperatures than a characteristic energy scale.
For instance, in a limit $N=1$, 
the system is reduced to a single Anderson impurity, 
and properties below the Kondo temperature 
are described by the Fermi-liquid theory.
\cite{YY1-2,ZlaticHorvatic,Nozieres}
Analogously, 
when the reservoirs are connected to a finite one-dimensional system 
of length $L$, a crossover seems to occur at a characteristic temperature 
$T_0 \equiv \hbar v_F/L$ ($v_F$ is the Fermi velocity) 
as pointed out by Kane and Fisher.\cite{KaneFisher}
They have discussed that a Fermi-liquid behavior is expected at $T \ll T_0$ 
due to the cut off by the length $L$,
and a Tomonaga-Luttinger behavior is expected at $T \gg T_0$. 
In our case, the length is $L = N a$ with $a$ being the lattice constant.
Since the characteristic temperature $T_0$ is a decreasing function of $N$,
the low-temperature region described by a Fermi-liquid 
becomes small for large systems.
Nevertheless, for small systems, 
the low-temperature region becomes relatively large.

In summary, we have studied the dc conductance 
through a finite Hubbard chain connected to noninteracting leads 
based on the perturbation theory. 
The application of the quasi-particle description used  
in the present study is not restricted to the single-mode case, 
and will be applied to various systems 
which have a Fermi-liquid ground state.

We would like to thank 
H. Ishii for valuable discussions.

\begin{table}
\caption{ Explicit expressions of $S_{2m+1}$ 
\label{table1}}
\begin{tabular}{rr}
$m$ &  $S_{2m+1}$   \\
\tableline
$  0$& $ 6 - \sqrt 3 \pi $ 
\\
$  1$& $ - 14/5 +  \sqrt 3 \pi/2 $ 
\\
$  2$& $ 290/147 -  5\sqrt 3 \pi/14 $ 
\\
$  3$& $ - 164998/105105 +   2\sqrt 3 \pi/7 $ 
\\
$  4$& $ 2797322/2111655 -      22\sqrt 3 \pi/91 $ 
\\
$  5$& $ - 6632866/5731635 +      11\sqrt 3 \pi/52 $ 
\\
\end{tabular}
\end{table}

\begin{figure}
\caption{ Schematic picture of the model: 
($\bullet$) interacting region, ($\circ$) noninteracting leads.} 
\label{fig:lattice}
\end{figure}

\begin{figure}
\caption{ Feynman diagram for $\,\Sigma^{(2)}_{jj'}(z)$.}
\label{fig:diagram}
\end{figure}

\begin{figure}
\caption{ 
$\mbox{Re} \Sigma_{jj'}^{(2)}(0) $ for odd $|j-j'|$ \ ($=2m+1$),
normalized as $\mbox{Re}\, \Sigma^{(2)}_{2m+1} (0)   
  = -t \left\{\rho(0) U \right\}^2  S_{2m+1}$  
with $\rho(0)=1/(2\pi t)$.
The inset shows the region $2\leq m \leq 11$.
Note that $\mbox{Re}\, \Sigma^{(2)}_{2m} (0) \equiv 0$.  
}
\label{fig:self2}
\end{figure}

\begin{figure}
\caption{ 
Size dependence of the conductance, $g_N$,   
for $\rho(0)U=1.0$. 
}
\label{fig:g_N}
\end{figure}

\begin{figure}
\caption{ 
Conductance for even $N$ ($=2M$), 
$g_{2M}$ vs.\  $M$, 
for several values of $\rho(0)U$. 
}
\label{fig:g_2M}
\end{figure}

\begin{figure}
\caption{ 
$U$ dependence of $g_N$ for large even $N$ ($=50$).
}
\label{fig:g_50}
\end{figure}

\end{document}